\pdfoutput=1
\documentclass[twocolumn,prb,showpacs,aps,superscriptaddress]{revtex4-1}

\usepackage[english]{babel}
\usepackage[ansinew]{inputenc}
\usepackage{times}
\usepackage{graphicx}
\usepackage{graphics}
\usepackage{amsmath}
\usepackage{amsfonts}
\usepackage{amssymb}
\usepackage{amsbsy}
\usepackage{array}
\usepackage{epstopdf}
\usepackage{makeidx}
\usepackage{color}
\usepackage{pgf}
\usepackage{bm}
\usepackage{tikz} 
\usepackage[normalem]{ulem}
\usepackage{multirow,bigstrut}
\usepackage{float}

\newcommand{\be}{\begin{equation}}
\newcommand{\ee}{\end{equation}}
\newcommand{\bea}{\begin{eqnarray}}
\newcommand{\eea}{\end{eqnarray}}

\makeindex
\begin{document}

\title{Dynamic Formation of Preferentially Lattice Oriented, Self Trapped Hydrogen Clusters}

\author{M. A. Cusentino}
\affiliation{Material, Physical, and Chemical Science Center, Sandia National Laboratories, Albuquerque, New Mexico 87185, USA}
\author{E. L. Sikorski}
\affiliation{Center for Computing Research, Sandia National Laboratories, Albuquerque, New Mexico 87185, USA}
\author{M. J. McCarthy}
\affiliation{Center for Computing Research, Sandia National Laboratories, Albuquerque, New Mexico 87185, USA}
\author{A. P. Thompson}
\affiliation{Center for Computing Research, Sandia National Laboratories, Albuquerque, New Mexico 87185, USA}
\author{M. A. Wood}
\affiliation{Center for Computing Research, Sandia National Laboratories, Albuquerque, New Mexico 87185, USA}

\date{\today}

\begin{abstract}
A series of MD and DFT simulations were performed to investigate hydrogen self-clustering and retention in tungsten.
Using a newly develop machine learned interatomic potential, spontaneous formation of hydrogen platelets was observed after implanting low-energy hydrogen into tungsten at high fluxes and temperatures.
The platelets formed along low miller index orientations and neighboring tetrahedral and octahedral sites and could grow to over 50 atoms in size.
High temperatures above 600 K and high hydrogen concentrations were needed to observe significant platelet formation.
A critical platelet size of six hydrogen atoms was needed for long term stability. 
Platelets smaller than this were found to be thermally unstable within a few nanoseconds.
To verify these observations, characteristic platelets from the MD simulations were simulated using large-scale DFT.
DFT corroborated the MD results in that large platelets were also found to be dynamically stable for five or more hydrogen atoms.
The LDOS from the DFT simulated platelets indicated that hydrogen atoms, particularly at the periphery of the platelet, were found to be at least as stable as hydrogen atoms in bulk tungsten.
In addition, electrons were found to be localized around hydrogen atoms in the platelet itself and that hydrogen atoms up to 4.2 \AA{} away within the platelet were found to be bonded suggesting that the hydrogen atoms are interacting across longer distances than previously suggested.
These results reveal a self-clustering mechanisms for hydrogen within tungsten in the absence of radiation induced or microstructural defects that could be a precursor to blistering and potentially explain the experimentally observed high hydrogen retention particularly in the near surface region.
\end{abstract}


\maketitle

\section{\label{intro}Introduction}
Future fusion reactors will require materials that can withstand the harsh conditions at the plasma-material interface including heat loads on the order of 10 MW/m$^3$ and particle fluxes on the order of 10$^{24}$ m$^{-2}$s$^{-1}$\cite{federici2001}.
In particular, the variety of plasma species, including the hydrogen fuel, helium ash, and impurities like beryllium or nitrogen, can damage the plasma-facing components resulting in microstructural changes and degradation of material properties.
The current candidate material for the divertor region, which will be subject to most of the particle exhaust from the core plasma, is tungsten due to its high melting temperature and thermal conductivity as well as its low sputtering yield\cite{pitts2013}.
A majority of the plasma will be comprised of deuterium and tritium where accumulation in the divertor can lead to poor material performance such as hydrogen induced embrittlement\cite{wirtz2015}. 
Additionally, due to the radiological hazards of tritium, there will also be limits on the amount of tritium that can be retained in the plasma-facing components, for example ITER will have a 1 kg inventory limit.
Therefore a detailed understanding of how hydrogen implantation and accumulation at trap sites into the tungsten divertor will affect material performance is critically needed.

Hydrogen is known to strongly trap at defects sites, like grain boundaries \cite{causey2002,zhou2010,jia_2016}, dislocations\cite{terentyev2014,bakaev2017}, and helium bubbles \cite{ueda2009,juslin2013_hydrogen,bergstrom2017}.
Relative trapping affinities at defects or within the bulk is key to understanding the damage accumulation; the former trap sites can be modified via manufacturing processes but the latter is an intrinsic property of the lattice.
Trapping at grain boundaries has been hypothesized as nucleation sites for bubbles and subsequent blister formation \cite{roth2011}.
However, this would only explain blister formation along grains and not within the grains themselves, where experiments have shown blister formation in single crystal tungsten \cite{enmoto2009}.
While hydrogen has a low solubility in tungsten, experiments have shown that tungsten exposed to hydrogen plasmas indeed result in large, micron sized blisters that form near the surface\cite{venhaus2001,ye2003,shu2007,balden2011}.
These blisters not only modify the near surface microstructure, including exfoliation and cracking\cite{ueda2005,shu2007}, but are a source of larger than expected hydrogen inventories in the divertor material.
Some prior work has suggested the formation of a supersaturated layer of tungsten, where the atomic percent of hydrogen is much higher than expected given the solubility in tungsten, at higher hydrogen fluences that may lead to blister formation \cite{alimov2005,enmoto2009,tanabe2014,li2023}
Similarly, irradiations of both ions and deuterium have shown higher retention due to trapping at ion-induced point defects like vacancies and voids\cite{hatano2013} and can remain trapped up to high temperatures.
However, tungsten exposed to pure deuterium plasmas at low energies, where the implantation energy is below the threshold energy for Frenkel pair formation, has still resulted in blister formation\cite{venhaus2001,ye2003}.
Without a continuous source for H trapping defects, there must be additional mechanisms at play to result in these observed high retention rates and super saturated regions leading to blistering.

Recent density functional theory (DFT) work has suggested that hydrogen can cluster into 2D platelet geometries in the absence of defects.
Li \textit{et al.}\cite{yang2019} indicated that stable hydrogen platelets can form below tungsten surfaces and is particularly stable under the (111) surface orientation.
Similarly, Hou \textit{et al.}, found hydrogen platelets on the (100) plane to be stable at high hydrogen concentrations and could be a mechanism for accumulation within the divertor \cite{hou2018}.
However, these prior DFT studies were performed for limited temperatures and simulation times making it difficult to dynamically  assess the stability of these types of platelet structures.
Simulations beyond DFT, such as molecular dynamics (MD), are needed to further determine the dynamics of platelet formation and growth.
Prior MD simulations by Smirnov \textit{et al.} \cite{smirnov2018} observed platelet formation in tungsten either under external non-hydrostatic stresses or near dislocations at hydrogen concentrations of 0.3-1 atom percent.
These studies indicate that there is a relationship between platelet formation and lattice strain that warrants further study.
These studies indicate a self-trapping mechanism for hydrogen that could explain the eventual formation of blister formation under low-energy, high flux deuterium exposure.

In this work, we simulate hydrogen implantation in tungsten using a newly developed W-H machine learned interatomic potential detailing platelet formation that occurs spontaneously at high hydrogen concentrations.
The effects of surface orientation, temperature, and flux on platelet growth are quantified.
In addition, we report on unprecedented large-scale DFT simulations to unravel the mechanism for stability of these platelets at high temperature.
The following sections describe simulation methods, hydrogen implantation MD simulations, and DFT simulations that validate the MD work for hydrogen platelet formation.

\section{\label{model} Computational Methods}
A combination of simulation methods is required in order to properly address hydrogen self-trapping within tungsten.
To achieve experimentally comparable particle fluxes, efficient classical MD simulations that utilize an interatomic potential are needed.
Direct simulation of hydrogen implantation is crucial because it allows for an assessment of the timescale of self-trapping in relation to particle flux and operational temperature.
Confirmation of MD observations are then carried out with density functional theory, scaling to simulation sizes only accessible with high-performance computing resources to ensure clarity in conclusions drawn. 
\subsection{Molecular Dynamics}

Molecular dynamics simulations were performed with the Large-scale Atomic/Molecular Massively Parallel Simulator (LAMMPS)\cite{plimpton1995,thompson2022}.
The Spectral Neighbor Analysis Potential (SNAP) \cite{thompson2015} developed for modeling W-H plasma material interactions was used to describe all atomic interactions.
The details of this interatomic potential (IAP) require a lengthy discussion which will be provided in another publication.
In short, this machine learned IAP was trained against DFT properties of gaseous, solid-state and interfacial phases of W-H.
Validation of the potential was done by comparing the extent of interpolation/extrapolation of a set of MD simulations to training structures.
These validation tests include H implantation and solubility in BCC W, providing confidence in the use cases herein.

A 6.4 nm x 6.4 nm x 12 nm tungsten slab with periodic boundary conditions in the lateral (x and y) directions and a free surface with a variable surface orientation (z) direction was generated, see Figure \ref{implant_image} for example.  
The slab was first equilibrated to 1000 K using an isobaric ensemble where the pressure components in the lateral (in-plane of free surface) directions were set to 0 bar for a total of 50 ps.
After equilibration, hydrogen was implanted by randomly placing a hydrogen atom 10 \AA{} above the surface and giving the hydrogen atom a velocity equivalent to an energy of 75 eV directly towards the surface.
The simulation was evolved for 10 ps before implanting another hydrogen atoms resulting in a flux of 2.4 $\cdot$ 10$^{27}$ m$^{-2}$s$^{-1}$.
During the implantations, the micro-canonical ensemble was used with a variable timestep that was allowed to vary between 10$^{-4}$ fs and 0.5 fs andwas updated every 10 timesteps such that no atom moved more than 0.02 \AA{}  per timestep in order to preserve stable dynamics.
After each implantation, a Langevin thermostat\cite{schneider1978} was run at 1000 K for 100 timesteps to remove any excess heat from the system and maintain the temperature at 1000 K.
Opposite the hydrogen exposed surface the last 1nm of material has a fixed position in space to ensure there is no net recoil of the sample.
The simulation was evolved for 1500 ion implantations or a total of 15 ns.
Hydrogen retention, diffusion rates, and clustering metrics were collected where analysis and visualization was performed with OVITO \cite{stukowski2010}.
Simulations were performed for the (100), (110), and (111) surface orientations, temperatures of 300 K, 600 K, 800 K, and 1000 K, and for fluxes of 2.4 $\cdot$ 10$^{27}$ m$^{-2}$s$^{-1}$, 1.1 $\cdot$ 10$^{27}$ m$^{-2}$s$^{-1}$, 6.1 $\cdot$ 10$^{26}$ m$^{-2}$s$^{-1}$, and 4.0 $\cdot$ 10$^{26}$ m$^{-2}$s$^{-1}$ by changing the lateral dimensions between 6.4 nm, 9.6 nm, 12.7 nm, and 15.9 nm.

\subsection{Density Functional Theory}
DFT calculations were performed using the Vienna Ab initio Simulation Package (VASP)\cite{Kresse1996}. Spin-polarized generalized gradient (GGA) exchange-correlation functional was used with the Perdew$\mbox{-}$Burke$\mbox{-}$Ernzerhof (PBE)\cite{Perdew1996} formulation. Plane-wave basis sets were implemented utilizing projector-augmented wave (PAW) pseudo-potentials. DFT run in molecular dynamics (DFT-MD) simulations utilized a cutoff energy of 700 eV and were sampled at the gamma point. 
Fermi smearing was utilized with a smearing width of 0.2 eV. The precision was set to normal and the electronic self-consistent loop exit criterion was set to 10$^{-4}$ eV. Single self-consistent framework (SCF) calculations on an DFT-MD snapshot of a 128 W atom supercell with 6 H atoms after 2 ps at 1000 K demonstrated that non-spin-polarized treatment was converged with spin-polarized treatment to 0.023 meV. Non-spin-polarized treatment was then used for the production DFT calculations. DFT-MD simulations used an NVT ensemble and 1 fs timesteps. Temperature was controlled using the Nos\'e-Hoover thermostat and the Nos\'e mass was set to correspond to 40 time steps (SMASS=0). DFT-MD simulations utilized the standard H pseudopotential and the GW W pseudopotential with 14 valence electrons. For visualization purposes, the partial charge density and electron localization function calculations \cite{Silvi1994} used the standard W pseudopotential with 6 valence electrons. This allowed for visualization of charge localized to H without core charge localized to all W atoms.

\subsection{Cluster Characterization}

All cluster identification for both MD and DFT was conducted using the Cluster Analysis modifier in the OVITO software \cite{stukowski2010}, using a cluster cutoff value of 4.2 \AA{}  for MD and 4.2 or 4.5 Angstroms for DFT-MD. 
Tracking unique H clusters through time is needed to discern mechanisms and kinetics of platelet formation and growth.
Once a cluster of at least two hydrogen atoms is identified by OVITO, the cluster is assigned a unique tag which is used to track its center of mass and hydrogen atom IDs. 
Thermally stable clusters taken for subsequent analysis are those that retain $>60\%$ of their H and collectively diffuse $<2$ \AA.
Orientation of a cluster is determined by a least squares fit to a plane containing all constituent H atoms.
The plane normal vector is then classified by comparing to set of low miller indexed planes of the W lattice (\{100\}, \{110\}, \{111\}, \{211\}).
Further comparison of H clusters based on tetrahedral and octahedral interstitial positions of the BCC lattice was performed if no low-index plane could be matched within $5^{\circ}$ projection difference.
Note that spherical clusters will fail to be classified within these tolerances as either crystal plane or interstitial types, an \textit{other} category is reserved for this case.
An added benefit of assigning persistent unique cluster IDs is to quantify the lifetime of any stable cluster, this includes both the growth and breakup modifications to a cluster.
Generalizable code for this analysis is available upon request.

\section{\label{md}Dynamics of Hydrogen Retention}
\begin{figure}[t]
\centering
\includegraphics[height=8.5cm]{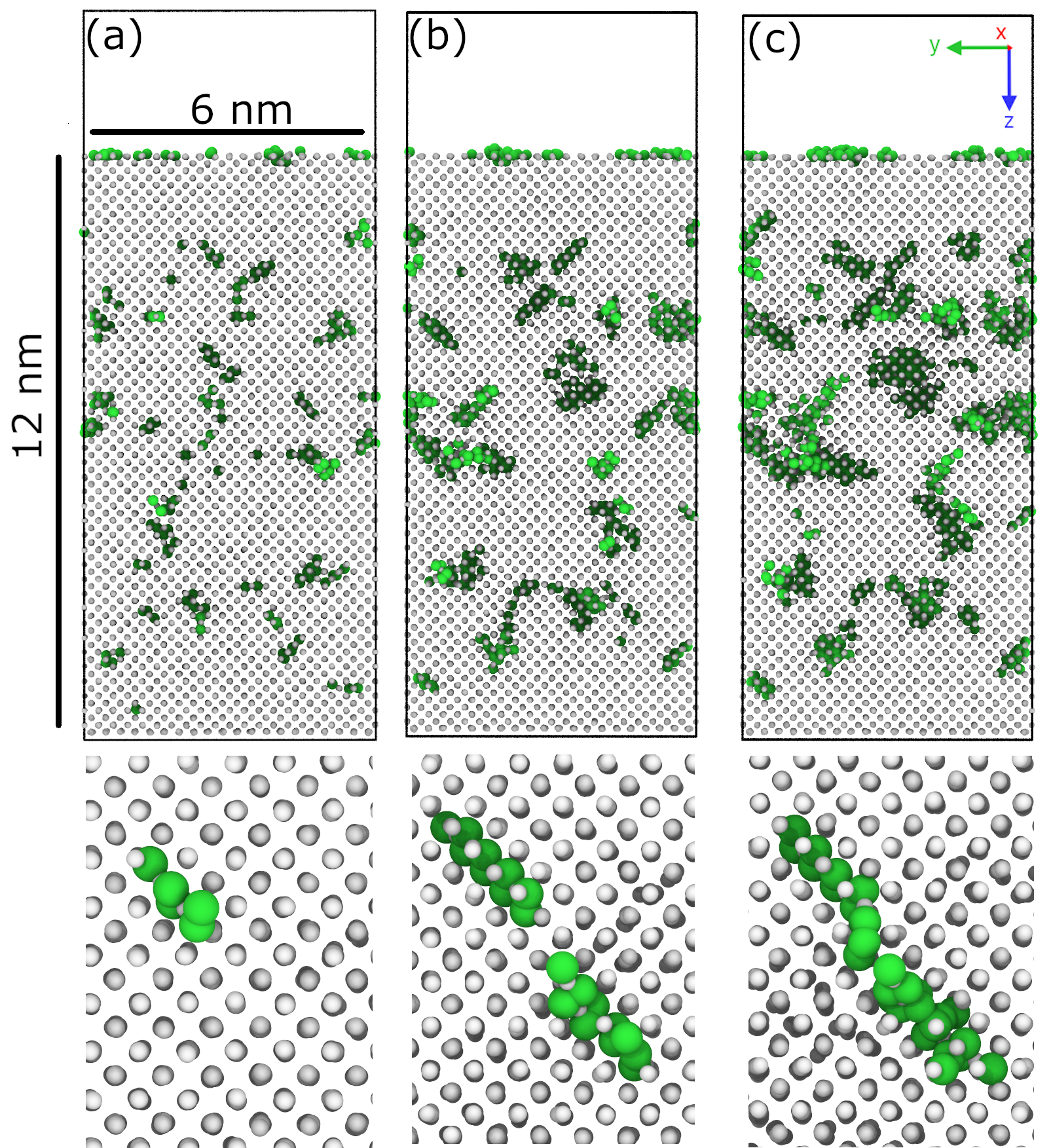}
\caption{\label{implant_image} Atomistic snapshots of 75 eV hydrogen implantation in tungsten at (a) 5 ns, (b) 10 ns, and (c) 15 ns.  The top image displays the entire simulation cell while the bottom image depicts a particular platelet growing over time.  The grey and green atoms represent tungsten and hydrogen, respectively.  
}
\end{figure}

In order to unravel the mechanisms for damage in the near surface region, detailing the dynamic changes in hydrogen accumulation is required.
Predictions of blister formation due to mass transport of the plasma-borne species within the matrix cannot be made from binary collision simulations of single ion strikes\cite{ziegler2008srim}.
Compounding effects of hydrogen accumulation are hypothesized to be sensitive to surface orientation, diffusion barriers and flux of particles, necessitating a direct investigation with MD simulations.
Herein we address each of these contributing factors in turn, ruling out weakly dependent variables to limit the number of redundant simulations.
If hydrogen was randomly distributed in the lattice then damage due to clustering (and thus local expansion of the lattice) will be depend on the energetic favorability, and diffusive processes that bring hydrogen into close proximity.
Key variables for study are then those that challenge this base assumption of random spatial distribution, energetic stability, and diffusive versus ballistic transport of hydrogen. 
Therefore, we prepare simulations that contrast i) surface orientation to address channeling along crystallographic directions, ii) diffusion driven transport at various temperatures, and iii) particle flux values that control ballistic transport through the lattice. 

We ran implantations in a representative selection of low energy surface orientations to highlight effects of energetic particle channeling and surface dependencies.
However, we did not observe drastic differences between the (100), (110), and (111) surface orientations.
For 75 eV particles it was observed that all orientations show significant adsorption at the surface, a depletion zone up to  1 nm below the surface, and a monotonically increasing accumulation up to a maximum depth of 12 nm.
Beyond fluences of 1.2 $\cdot 10^{19}$ m$^{-2}$ ($>5$ns) these profiles are self-similar for all surface profile, with very little variation between surface orientations.
For the hydrogen retention, the (100) and (111) surfaces had similar hydrogen retention at  47.1\% and 47.9\%, respectively.
In contrast, the (110) surface had a significantly lower retention at 28.3\% which was in part due to the higher reflection rate.
The percentage of hydrogen at the surface ranges from about 5\% for the (111) surface to about 8\% for the (110) surface.

\begin{figure}[t]
\centering
\includegraphics[width=8cm]{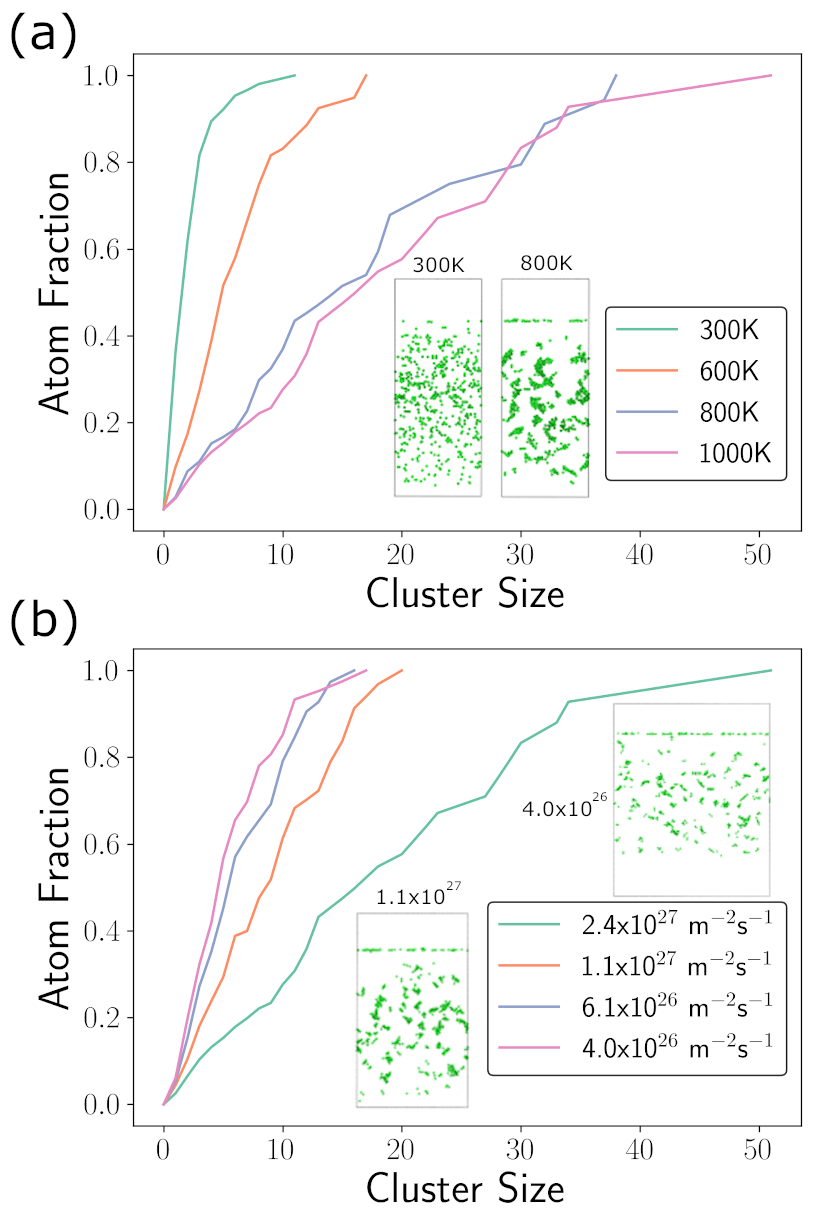}
\caption{\label{parameters} Cumulative distribution function of hydrogen cluster sizes after 15 ns of simulation time for varying (a) temperatures, and (b) hydrogen fluxes. Inset images in each panel visualize the simulation cell highlighting cluster morphology. 
}
\end{figure}

Throughout the simulated implantations, prompt formation of hydrogen clusters was observed for all plasma-material exposure conditions. 
Direct simulation of the implantation and time evolution of absorbed plasma species allows for diffusion and self-trapping events to occur naturally, without the need to bias towards clustering phenomena. 
Snapshots in Figure \ref{implant_image} are at 5 ns intervals are shown for the (100) surface at 1000 K and a flux of $2.8 \cdot 10^{27} m^{-2}s^{-1}$.
Within the first 2 ns there is little hydrogen clustering, but mobility is observed to be high throughout the simulation cell.
Weak dependence on the surface orientation for cluster size (data captured at 15ns elapsed) was observed, with the (110) surface orientation resulting in smaller clusters due to lower overall retention.
This diffusive growth of hydrogen clusters is directly amplified at elevated temperatures, as exemplified by Figure \ref{parameters}, where all temperature changes were run with a surface normal of (100) and $2.8 \cdot 10^{27}$ $m^{-2}s^{-1}$ flux.
Interestingly, the cluster size distribution did not increase further moving from 600 K to 1000 K.
This implies that the increase in diffusivity needed to form larger clusters is beginning to be offset by the decreasing thermal stability of additional hydrogen.

Due to limitations on length and timescale available to MD, particle fluxes are higher than what is expected at the ITER diverter surface. 
At the implantation energy studied here a high concentration of absorbed hydrogen resides close to the surface, promoting clustering via short diffusive distances needed and ballistic implantation into preexisting clusters.
Indeed, the largest cluster sizes ($>50$ atoms) were observed for the highest fluxes tested, where a flux of six times lower resulted in mean cluster sizes a factor of four lower.
These observations are captured in Figure \ref{parameters}, where insets demonstrate differences in the final state over the range of flux values tested, but where the total hydrogen content is equivalent.
The drastic change in cluster size distribution observed between fluxes of $2.4\cdot10^{27}$ and $1.1\cdot10^{27}$ is evidence of the increase in H self-trapping events where mobile H in the lattice will attach and grow larger clusters. 

Simulations of hydrogen clustering over material and environmental conditions reinforces the importance of understanding the mechanisms of divertor surface degradation as all conditions predict a rapid change in local chemistry and lattice strain.

\begin{figure}[t]
\centering
\includegraphics[width=8.5cm]{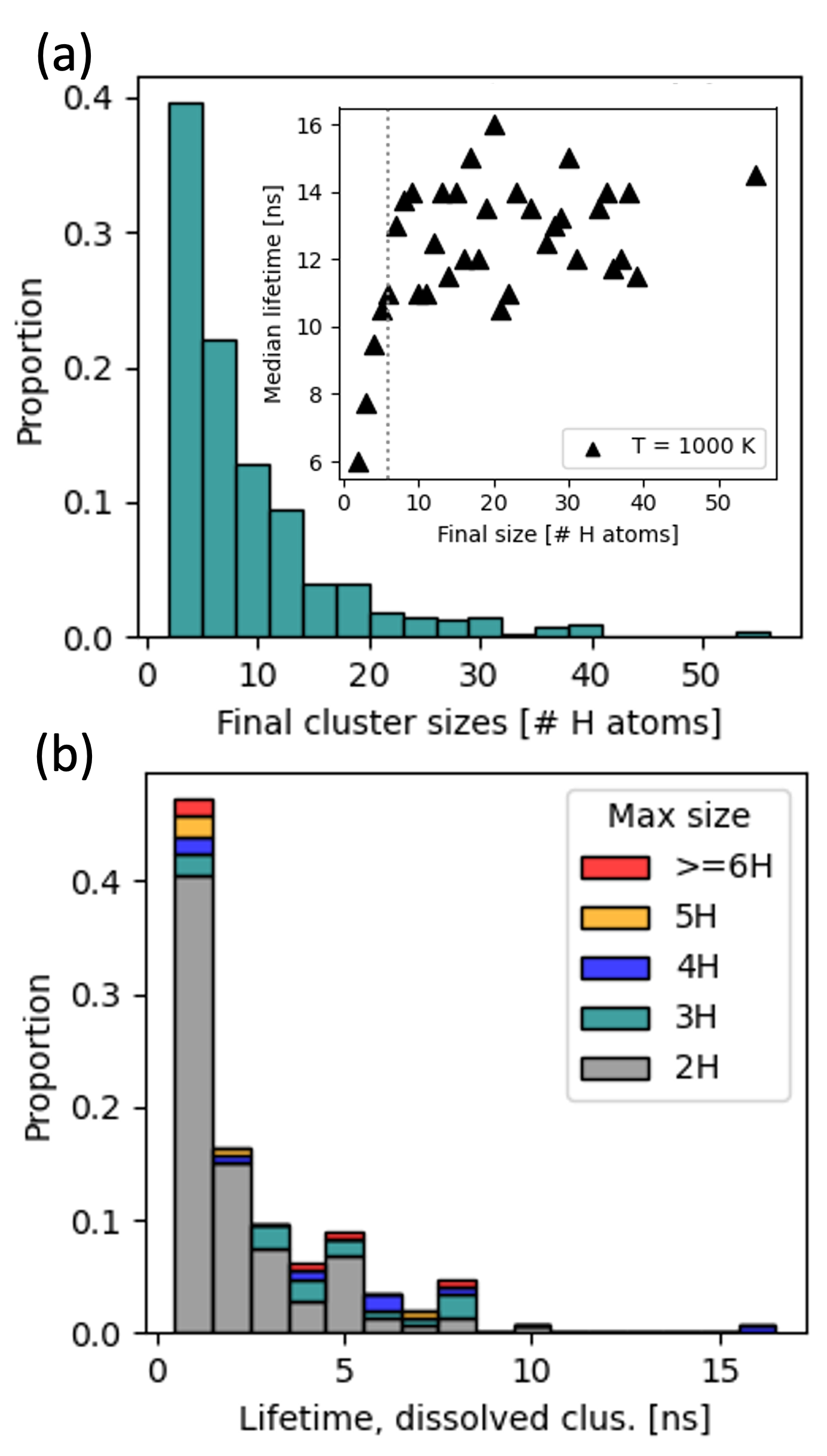}
\caption{\label{lifetime} (a) Final sizes for hydrogen clusters for simulations at T = 1000 K. Inset image captures median lifetime of each integer cluster sizes, where a clear increase in stability is seen for larger cluster sizes.
(b) Distribution of lifetimes for dissolved clusters (i.e., those that disappear before the simulation end time). 
The colored sections of each bar indicate the maximum size reached by dissolved clusters, which only extremely rarely exceeds seven hydrogen atoms.   
}
\end{figure}

The most surprising observation was the interplay of the lattice and hydrogen cluster orientations. 
Classical arguments for the formation of a secondary phase in a solid would be that free energy is minimized were the surface area between phases is reduced, implying  spheroidal shapes were to be expected.
However, as shown in Figure \ref{implant_image}, larger clusters atoms tend to form rather flat, distinctly oriented structures. 
Stability of these larger structures hinges upon whether they can trap enough free hydrogen to reach a critical size at which they become immobile and effective sinks for additional hydrogen. 
This critical size for stability would constitute an initial precursor to blister formation. Below that size, hydrogen has the potential to desorb from the cluster and freely diffuse through the tungsten matrix.

Analysis of these mechanisms of blister formation was conducted using the cluster tracking techniques outlined in Section II C. 
Figure \ref{lifetime}(a) and (b) captures four simulations run at 1000 K, with surfaces of (100) and all fluxes tested (between $0.4$ to $2.8 \cdot 10^{27}$ $m^{-2}s^{-1}$, see also Figure \ref{parameters}b). 
Data is shown at the end of the simulated implantation, 16 ns, where more than half of the observed clusters contain five or more hydrogen. 
The inset shows the median lifetime per cluster size for the same data. 
Clusters with sizes smaller than approximately 6 hydrogen (dotted grey line) have significantly shorter lifetimes than those with more than 6 hydrogen, which all have lifetimes of at least 11 ns. 
Notable as well is the steep increase in median lifetime for clusters from sizes 2 to 6, which demonstrates the strong contribution of even single hydrogen atoms to self-trapping effectiveness. 
Beyond the apparent threshold of 6 hydrogen atoms, adding hydrogen does not appear to either strengthen or weaken the dynamic stability of already-formed clusters.

\begin{figure}[t]
\centering
\includegraphics[width=8.5cm]{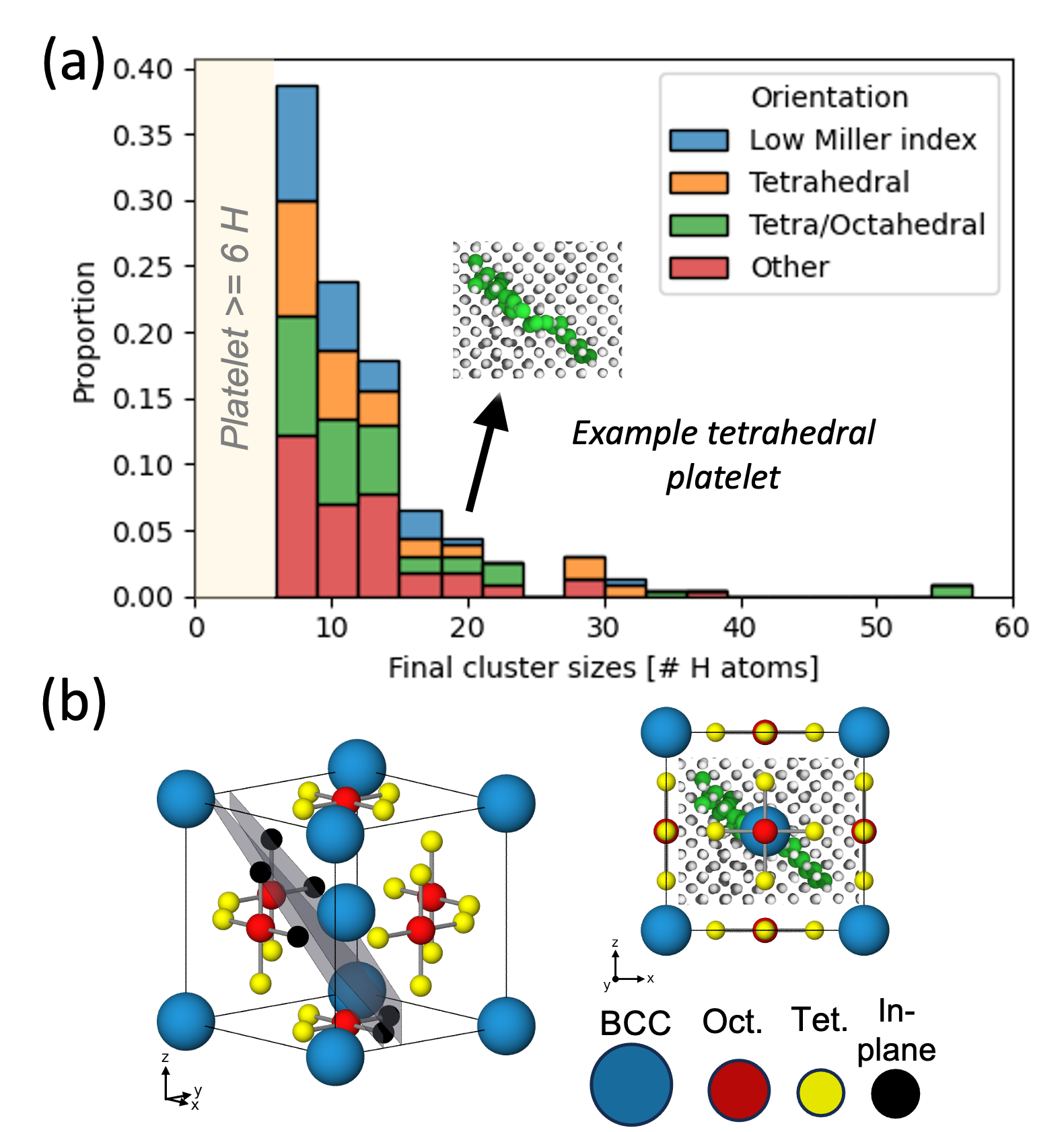}
\caption{\label{orientation} (a) Stable cluster orientation categorized by habit plane within the BCC crystal. Low Miller index category (blue) includes {100}, {110}, {111} and {211} planes. Purely tetrahedral occupation (orange) is isolated from mixed tetrahedral/octahedral (green) clusters. (b) Schematic of a conventional unit cell detailing the tungsten BCC lattice (blue atoms), tetrahedral interstitial sites (yellow), and octahedral interstitial sites (red). The grey lines connecting the tetrahedral and octahedral sites on each face of the cell are added to guide the eye. 
}
\end{figure}

Tracking the cluster lifetimes of sub-critical sizes, Figure \ref{lifetime}b) highlights the rapid ($<1 ns$) breakup of small clusters implied by Figure \ref{lifetime}b).
For comparison, the total number of large ($\ge 6$ H) clusters seen to breakup was 5 and is shown in red.
These findings support the existence of a stability threshold of approximately six hydrogen atoms to maintain and grow large hydrogen clusters.
To ensure this stability threshold is not an artifact of the interatomic potential, we will turn to first principles predictions of hydrogen clustering in Section IV.

As alluded to previously, and shown in Figure \ref{implant_image}, larger hydrogen clusters inside of the tungsten bulk tend to form plate-like, oriented structures where the thickness is most commonly one H layer. 
The specific orientation and geometry of these large clusters is important to quantified as the lattice strain generated by these self-trapping loci will lead to damage accumulation and increased hydrogen inventory. 
Orientation analysis will be conducted on the same simulations as those  in Figure \ref{lifetime} (namely, those at T = 1000 K).
Prior simulation work\cite{hou2018} has indicated that clusters of hydrogen can form stably along the (100) plane, but crucially these predictions were made at 0 K, neglecting thermal expansion and dissociation. 
Figure \ref{orientation}c) highlights all of the available tetrahedral and octahedral sites within a single unit cell of the W-BCC lattice. 
Additionally, an example plane that connects the energetically favorable tetrahedral sites is shown in grey.
For hydrogen clusters that do not exhibit a clear planar orientation, see superimposed example in Figure \ref{orientation}d), the closest plane indices are determined within a small tolerance. 
The distribution of these determined cluster orientations is displayed in panel b) where low indexed planes and those formed by connecting adjacent interstitial sites far out number cluster with no dominant orientation. 
Data categorized as 'Tetrahedral' is exclusively where hydrogen resides in adjacent interstitial sites of this type, where a cluster of mixed interstitial sites is given its own category. 
Of the stable clusters these oriented platelets form readily, indicating that a maximization of H-H interaction is not the most energetically favorable configuration. 
This understanding of preferential sites within the lattice during platelet formation provides the ability to make miniaturized versions that can be investigated further with first principles methods. 
Namely, the nature of the bonding between W-H and H-H is best characterized with these higher fidelity methods.


\section{\label{dft} Mechanism for Stable Hydrogen Self Trapping}
The primary challenge for DFT calculations is the cubic increase in computational cost associated with respect to number of electrons.
To mitigate this, smaller representative problems have to be utilized, with the caveat that the predictions be converged with respect to finite size and time effects.
Presently, the main concern is that smaller periodic cells will artificially increase the hydrogen self interactions. 
An initial sweep of supercell sizes ranged from $2 \times 2 \times 2$ to $7 \times 7 \times 7$ containing 4 to 12 H atoms arranged in adjacent tetrahedral sites. Though the SCF per atom H incorporation energy for the $5 \times 5 \times 5$ supercell was converged with the $7 \times 7 \times 7$ supercell to 0.04 eV, we wanted to mitigate finite size effects and enable investigation of higher H atom counts as much as computationally feasible. We selected the $7 \times 7 \times 7$ supercell for further study.
It is important to note that smaller volumes were observed to suppress the formation of a thermally stable cluster, rather showing that spatially dispersed hydrogen is energetically favorable. 

\begin{figure}[htb]
    \centering
    \includegraphics[width=0.45\textwidth]{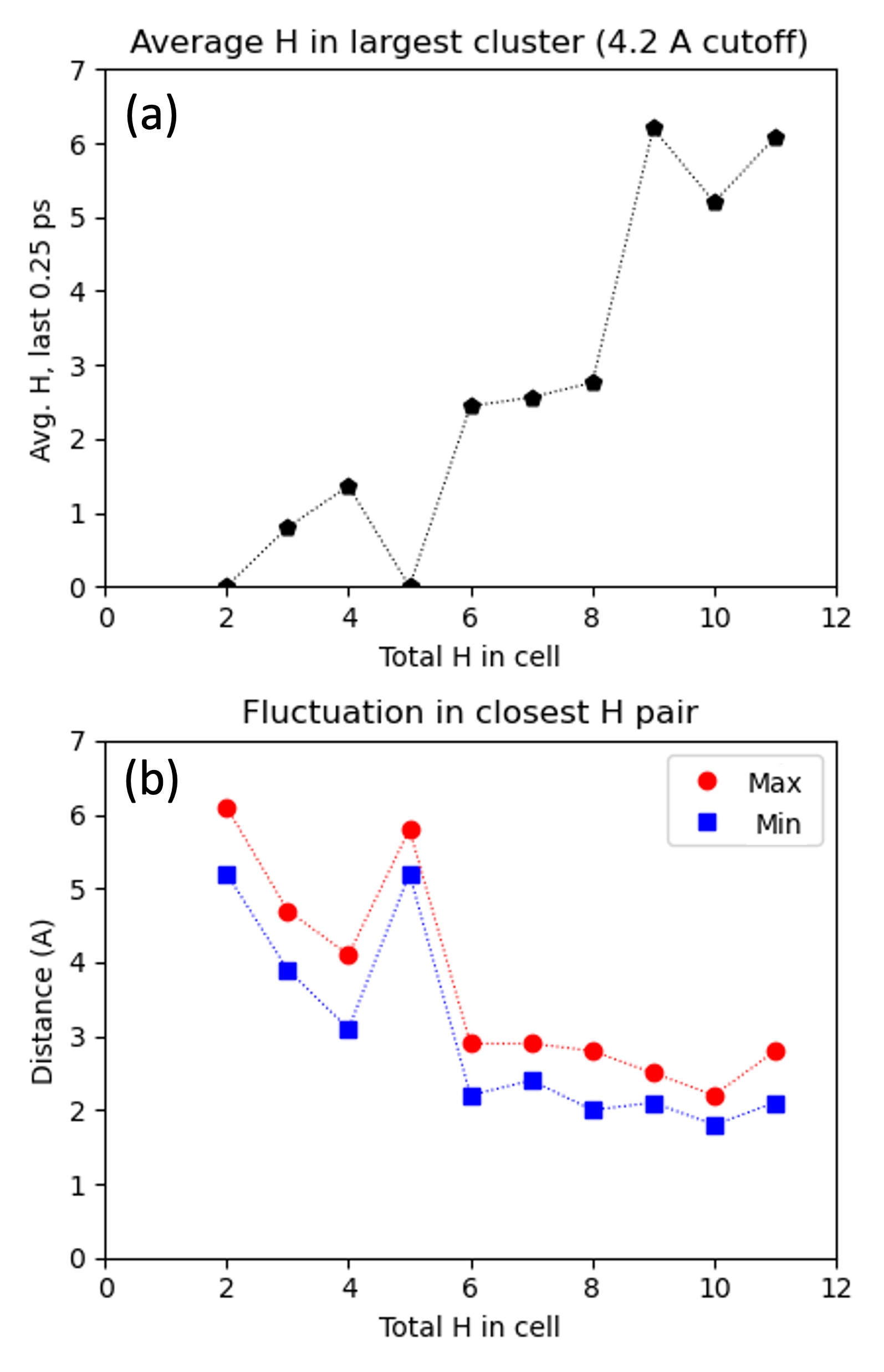}
    \caption{ Statistics tracking the dynamics of hydrogen atoms in a series of DFT-MD runs of a 7x7x7 W supercell (686 atoms) with increasing hydrogen concentration. (a) Average number of H atoms in the largest-tracked cluster. Each value is averaged over the last 0.25 ps of a 1 ps simulation. (b) Measurement of the fluctuations between the two closest hydrogen atoms in each simulation. The minimum (blue) and maximum (red) distances found over time are plotted. With increasing hydrogen concentration, both the minimum and maximum distances decrease significantly, corresponding to an increase in tracked cluster size. }
    \label{fig:AIMD-stats}
\end{figure}

To investigate what mechanisms may be bolstering platelet stability, the  $7 \times 7 \times 7$ supercells included a range of H atoms from 2 to 11 and each was simulated at 1000 K for 1 ps. We note these large DFT-MD simulations took $\approx$ 37k node hours on the GPU machine Summit at Oak Ridge National Laboratories.
The H atom range allowed for investigation of early stage platelet formation and comparison across concentrations. 
Using the same spatial cutoff as the large scale implantation simulations (4.2 \AA), the average number of clustered hydrogen was quantified, which is displayed in Figure \ref{fig:AIMD-stats}a). 
As the cluster values can shift somewhat through time, the average size of the largest detected cluster is recorded from 0.75 - 1.0 ps.
Figure \ref{fig:AIMD-stats}b) displays the average distance between the closest two H atoms in the simulation (usually located in the clusters from Figure \ref{fig:AIMD-stats}a).
Their average distances are decreased as the H concentration increases in the cell, pointing to an increased stability from H-H interactions.
In agreement with the classical MD predictions, there exists a minimum H count for stable clusters which occurs around 6-8 atoms based on these DFT-MD tests at 1000 K. 
The agreement between the MD and DFT-MD tests is quite remarkable, as no training data of this scale was included in the initial MLIAP training set.

\begin{figure}[!t]
    \centering
    \includegraphics[width=0.45\textwidth]{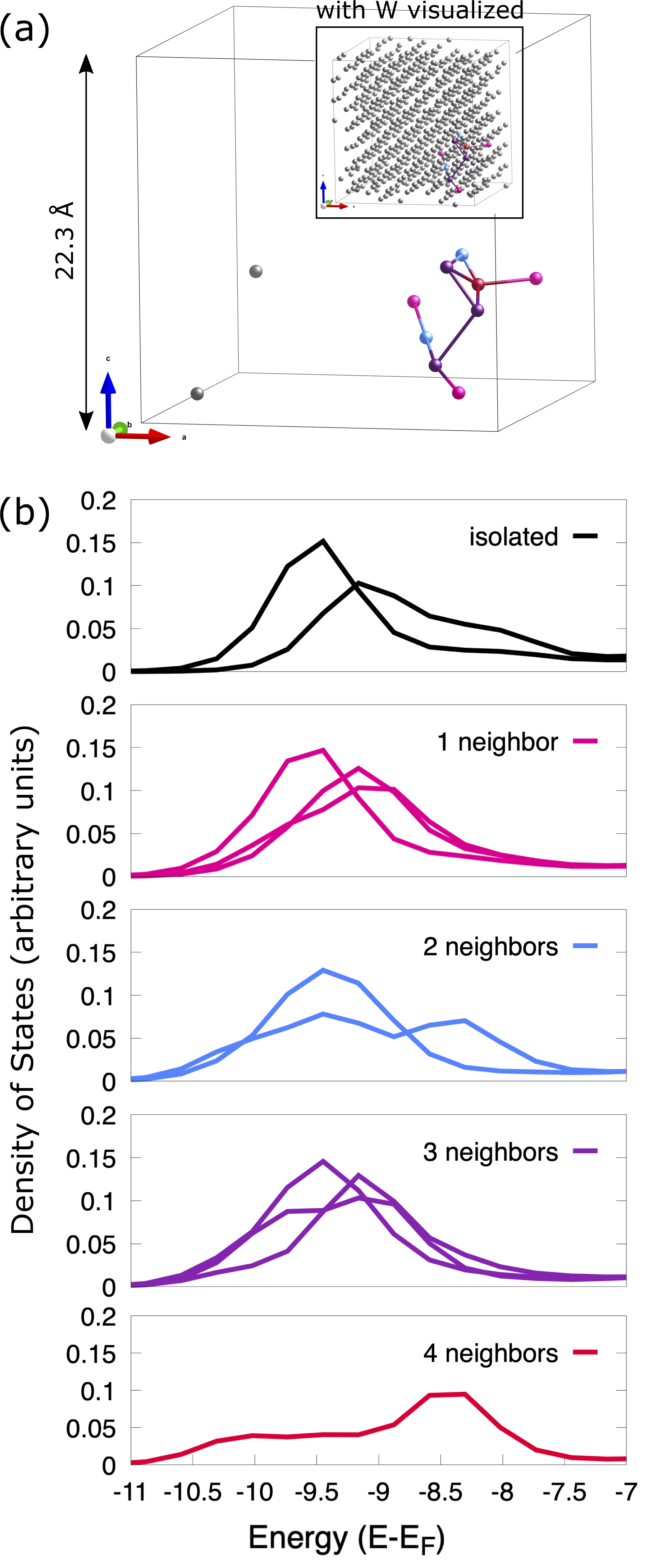}
    \caption{(a) Visualization of H atoms in the 11H DFT-MD simulation after $\approx$ 0.8 ps. W atoms are only displayed in the inset so as to better view the position of the H atoms. Grey, pink, blue, purple, and red represent H with 0, 1, 2, 3, and 4 H neighbors, respectively, with a 4.5 $\mbox{\AA}$ cutoff radius. Bonds are added to the H solely for visual aid. (b) Local density of states for each H atom with respect to the Fermi level.}
    \label{fig:DOS}
\end{figure}

\begin{figure*}[!ht]
    \centering
    \includegraphics[width=0.9\textwidth]{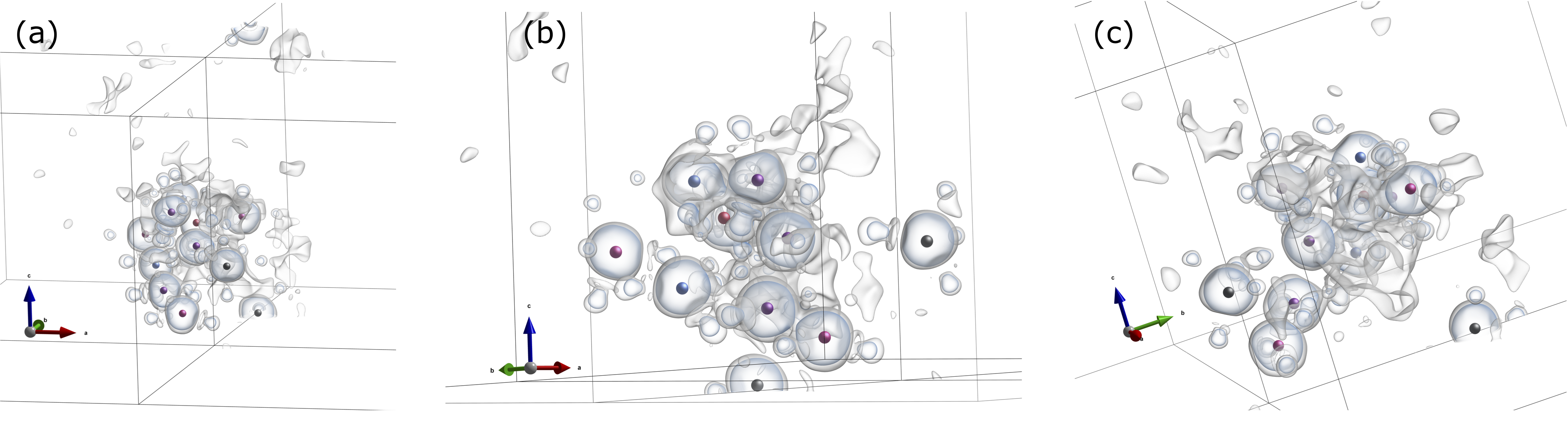}
    \caption{Partial charge density of W supercell with 11 H. Blue and grey isosurfaces correspond to $\approx$84\% and 56\%, respectively, of the total charge density from -11 to -6 eV below the Fermi level. H atoms are colored in agreement with Figure \ref{fig:DOS}. The simulation cell is shifted for clarity of shared partial charge among hydrogen.}
    \label{fig:PCD}
\end{figure*}

A snapshot from the highest hydrogen concentration after $\approx$ 0.8 ps was used to investigate the electronic structure leading to the bonding changes of interest. 
To better understand the electronic structure of H during platelet formation, we first look at the Local Density of States (LDOS). 
Any overlap between DOS curves is an indication of hybridization or bonding. 
The position of H in the W supercell in the DFT-MD snapshot are shown in Figure \ref{fig:DOS}(a). The H are color coded by the number of neighbors using a 4.5 $\mbox{\AA}$. H bonds are visualized at the same cutoff to better illustrate the distances and angles between the H atoms, but do not necessarily indicate real bonds. Examining atomic positions in Figure \ref{fig:DOS}(a), we can see all H atoms except two may be located in a platelet. 
We resolved the LDOS for every H in the supercell, and are displayed by neighbor count in Figure \ref{fig:DOS}(b).  
Peaks at lower energies and thus further away from the Fermi level are interpreted as more stable H, while higher energies are interpreted as less stable.
The LDOS peaks for H atoms ranging from 0 neighbors (isolated) to 3 neighbors indicate similar energetic stability. In contrast, the central H atom in the platelet has a much higher energy LDOS peak, indicating this H is less energetically stable. As additional H began to coalesce and take advantage of favorable platelet edge locations, the central H atom became trapped in the center of the platelet. Collectively, the LDOS suggests that H can exist favorably both in an isolated position and at the edges of a platelet. Having four or more neighbors is less energetically stable and this may indicate why platelets, with increased surface area, preferentially form over bubbles, with increased volume. As H concentrations increase, isolating may no longer be possible and thus the edge of a platelet becomes the most stable, possible configuration.

While overlap or hybridization in DOS peaks may suggest bonding, we need to check whether the electrons are also shared spatially. 
The partial charge density (PCD) for the supercell snapshot resolved from -11 to -6 eV below the Fermi level is shown in Figure \ref{fig:PCD}. 
The blue isosurface shows that each H has localized charge.  
Visualizing 56\% of the PCD illustrates longer range electron sharing. 
Both the blue and grey isosurfaces show the H atoms previously categorized as isolated do not share electrons with any other H atom. One of the H atoms categorized as having one neighbor (pink) does not in fact share charge, as seen on the left side of \ref{fig:PCD}(b). The remaining 8 H atoms are all contained in the same grey isosurface, shown in \ref{fig:PCD}(b) and \ref{fig:PCD}(c).

A spatial cutoff of 4.2 $\mbox{\AA}$ will identify 7 of the H atoms contained in the grey isosurface as part of the same cluster. 
However, a cutoff value of 4.5$\mbox{\AA}$ is needed to classify all 8 H as part of the cluster, but this incorrectly includes one of the isolated H as well which does not share charge with the cluster. 
While using a radial cutoff metric may not precisely identify all H atoms sharing charge, we can conservatively use a 4.2 $\mbox{\AA}$ cutoff distance to identify H platelets that share electrons. This is in good agreement with the findings from our MD simulations.

\begin{figure}[!t]
    \centering
    \includegraphics[width=0.45\textwidth]{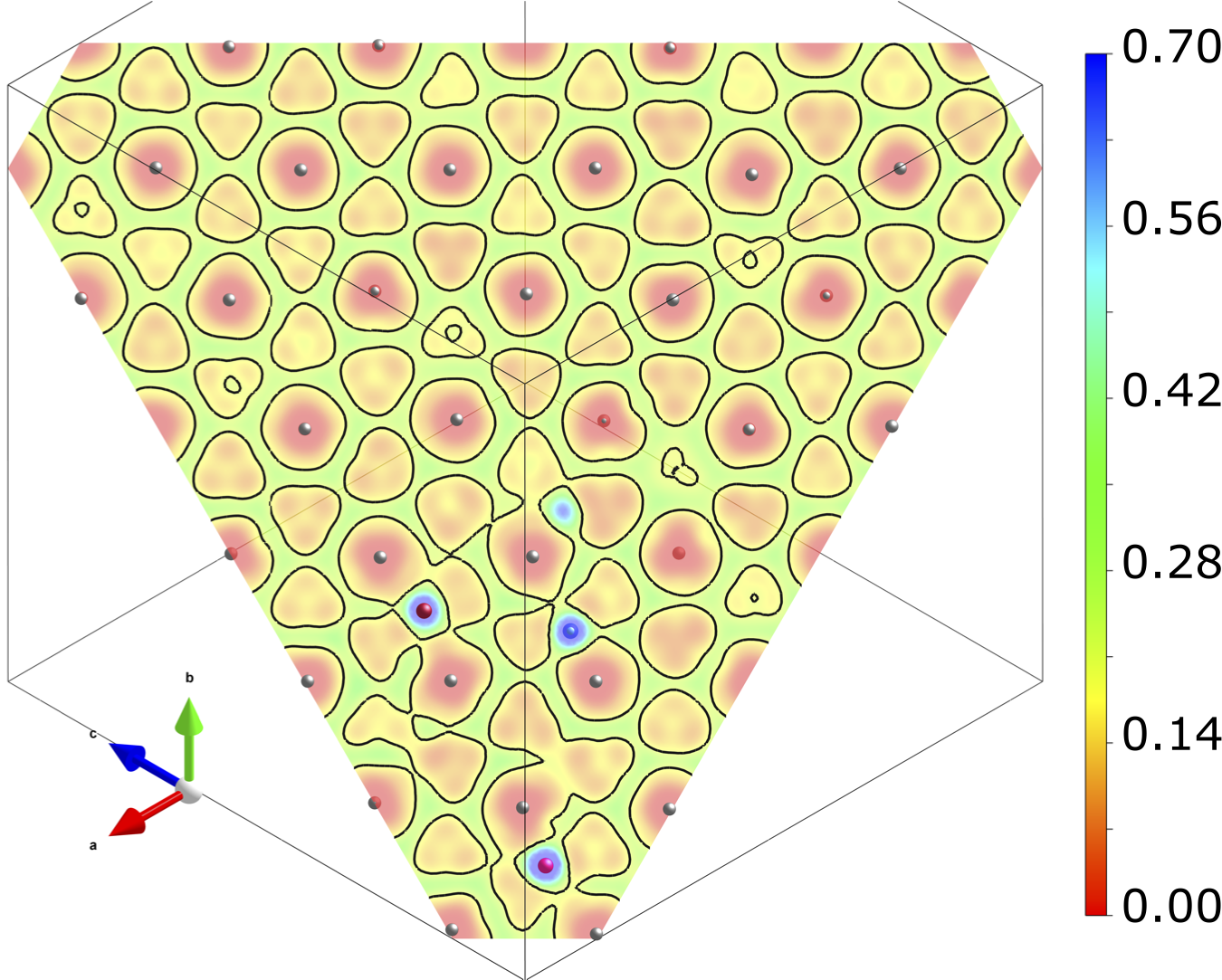}
    \caption{Electron Localization Function for ($11\overline{1}$) slice intersecting three of the H atoms. The ELF intersects the charge localized to one additional H that does not precisely lie on the plane. The ELF color range is narrowed to 0.0 to 0.7 for visualization. Blue represents a high probability of finding an electron and red represents a low probability of finding an electron. Contours are added as a visual aid.}
    \label{fig:ELF}
\end{figure}

While the PCD reveals an energy resolved charge map, we can get a complete spatial picture of the electronic structure by also examining the Electron Localization Function (ELF). 
The ELF maps the probability of finding any calculated electron from 0.0 to 1.0, where 0.0 indicates a low probability of finding an electron and 1.0 indicates a high probability of finding an electron.  
The ELF is scaled such that a value of 0.5 would match the probability of a homogeneous electron gas. 
The ELF for the 0.8 ps snapshot is shown in Figure \ref{fig:ELF}. 
It is visualized as a slice along ($11\overline{1}$) that intersects with 3 H atoms and captures some of the charge localized to a fourth H that does not precisely intersect the same plane. 
The ELF reveals that there is a very high (ELF $\geq$0.7) probability of finding an electron localized to each H in the slice. 
Meanwhile, we can also observe how the presence of H affects the surrounding W electron sea. 
On the upper portion of the ELF slice, we can see uniform green regions (ELF $\approx$ 0.3) depicting the electron sea between W atoms. 
In contrast, on the lower portion and near the Hs, the electron sea is discontinuous. 
Increased regions of yellow and orange (ELF $\approx $0.1 - 0.2) are present between W atoms near the H atoms. 
This suggests that the H atoms are stealing charge from the neighboring W atoms and creating these charge voids.

\section{\label{conclusions}Conclusions}
In this work, we have performed MD simulations of hydrogen implantation in tungsten using a newly developed W-H machine learned interatomic potential and ran large-scale DFT simulations of hydrogen clustering in unprecedentedly large simulation cells.
After low-energy, high flux hydrogen implantation in tungsten, we observe the spontaneous formation of 2D platelet hydrogen clusters which accumulate along preferred directions, especially those containing tetrahedral sites, such as the (111) plane.
The platelets require a high concentration of hydrogen and high temperatures above 600 K to regularly form and tend to become more stable as they become larger.
A critical size of about 6 hydrogen atoms was needed to form a stable platelet, where clusters smaller than this tend to annihilate within a few nanoseconds.
Characteristic platelets formed during the MD simulations were modeled using DFT and also found to be energetically stable and remained clustered at high temperatures.
Hydrogen atoms, particularly at the periphery, were found to be at least as stable as hydrogen atoms in the bulk as indicated by the LDOS.
In addition, ELF analysis showed that electrons had a high probability of being located near a hydrogen atom in a platelet and that the local electronic structure near the surrounding tungsten atoms was distorted, indicated that the hydrogen were stealing charge from tungsten to become more stable in the platelet.
The combination of MD and DFT simulations indicate that these platelets can form form under fusion relevant conditions and are stable at the relevant temperatures.

The results discussed in this work indicate a mechanism for hydrogen self-clustering at high temperatures and fluences expected in the divertor region.
Tungsten exposed to low-energy hydrogen has resulted in blistering yet the exact mechanism for blister formation is still unknown particularly for blisters that form within the grain.
This work suggests that hydrogen can cluster into platelets which may be a precursor to the supersaturated layer and blistering observed in experiments, which would require large inventories of hydrogen to accumulate in the material.
If hydrogen can self-cluster at higher concentrations and temperatures, retention and blistering will remain an issue even in the absence of neutron induced radiation damage or other microstructural defects that traditionally have shown to trap hydrogen.
This could potentially result in higher than expected hydrogen retention in the material just from the hydrogen implantation itself, which will affect the overall tritium inventory in the reactor.
Additional work is needed in extending the time and length scales of the observed platelet growth to mesoscale models, which do not currently take into account this mechanism, to bridge the gap between hydrogen platelet formation and blistering.

\begin{acknowledgments}
All authors gratefully acknowledge funding support from the U.S. Department of Energy, Office of Fusion Energy Sciences (OFES) under Field Work Proposal Number 20-023149 and the plasma surface interaction project of the Scientific Discovery through Advanced Computing (SciDAC) program, which is jointly sponsored by the Fusion Energy Sciences (FES) and the Advanced Scientific Computing Research (ASCR) programs within the U.S. Department of Energy Office of Science. This research used resources of the Oak Ridge Leadership Computing Facility, which is a DOE Office of Science User Facility supported under Contract DE-AC05- 00OR22725.

This article has been authored by an employee of National Technology \& Engineering Solutions of Sandia, LLC under Contract No. DE-NA0003525 with the U.S. Department of Energy (DOE).
The employee owns all right, title and interest in and to the article and is solely responsible for its contents. 
The United States Government retains and the publisher, by accepting the article for publication, acknowledges that the United States Government retains a non-exclusive, paid-up, irrevocable, world-wide license to publish or reproduce the published form of this article or allow others to do so, for United States Government purposes. 
The DOE will provide public access to these results of federally sponsored research in accordance with the DOE Public Access Plan https://www.energy.gov/downloads/doe-public-access-plan. 
This paper describes objective technical results and analysis. Any subjective views or opinions that might be expressed in the paper do not necessarily represent the views of the U.S. Department of Energy or the United States Government.
\end{acknowledgments}

\section*{References}
\bibliography{wh}
\bibliographystyle{unsrt}

\end{document}